\documentclass{jltp}

\usepackage{graphicx} 
\title{A temperature-controlled device
for volumetric measurements of Helium adsorption in porous media.}

\author{B. Cross, L. Puech, P. E. Wolf}

\address{Institut N\'eel, CNRS-UJF, BP 166,
38042 Grenoble-Cedex 9, France\\
}

\runninghead{B. Cross, L. Puech, P. E. Wolf}{A temperature-controlled device
for measurements of Helium adsorption}

\begin{document}

\maketitle

\begin{abstract}
We describe a set-up for studying adsorption of helium in silica
aerogels, where the adsorbed amount is easily and precisely controlled
by varying the temperature of a gas reservoir between 80~K and 180~K.
We present validation experiments and a first application to aerogels.
This device is well adapted to study hysteresis, relaxation, and
metastable states in the adsorption and desorption of fluids in porous
media.

% Adsorption
% 
% fluid surfaces, 68.03.Ðg
% 
% quantum fluids, 67.70.+n
% 
% solid surfaces, 68.43.Ðh
% 
%64.60.-i 	General studies of phase transitions (see also 63.70.+h Statistical mechanics of lattice vibratio
% 64.70.Fx 	Liquid-vapor transitions
% 67.70.+n 	Films (including physical adsorption)
%68.03.Cd 	Surface tension and related phenomena
\end{abstract}

\section{INTRODUCTION}\label{intro}
The technique of adsorption isotherms, which measures the adsorbed
amount of a fluid in a porous medium as a function of the applied
pressure, is a central tool in surface physics.  On the applied side,
isotherms are used to characterize the porous medium itself
(distribution of pores sizes), while, on the fundamental side, they
detect the phase transitions of the confined fluid
\cite{Gelb99}.  Adsorption of helium has been studied in a variety of
materials with different topologies, such as grafoil, carbon
nanotubes, or silica glasses.  The case of silica aerogels has
attracted attention in the recent years, since the early work of Wong
and Chan \cite{Wong90}.  As in other porous media, experiments show
that the isotherms are hysteretic between adsorption and desorption,
and that condensation occurs over a range of pressures
\cite{Tulimieri99,Gabay00,HermanTout,Lambert04}.  However, this range
is unexpectedly narrow, compared to what could be expected from the
wide distribution of "pores" sizes in aerogels, especially at low
porosity \cite{Tulimieri99,HermanPRB05} or low temperature
\cite{Lambert04,HermanPRB05}.  Both facts are in agreement with recent
theoretical work \cite{Kierlik01,Detcheverry05}, according to which
the shape of the isotherm, and its hysteresis, result from the
disordered nature of the aerogel, rather than of its geometry or
topology.  In particular, the observed change of shape with porosity
or temperature could sign the occurrence of a disorder induced,
out-of-equilibrium, phase transition\cite{Kierlik01}.  While this
vision is supported by optical experiments \cite{Lambert}, a challenge
would be to study in detail the critical shape of the isotherm close
to the transition.  An other open problem in the context of adsorption
is whether this hysteresis disappears at, or below, the bulk critical
point.

Studying these questions require high resolution measurements of the
isotherms.  Usually, the adsorption curves are determined by a
volumetric technique, by adding known doses of helium in successive
steps and measuring the resulting equilibrium pressure after each
step.  This has several drawbacks.  If large steps are used, the flow
rate at which helium is injected is initially large, and this warms up
the sample, due to the heat released by the adsorption process.  This
might perturb the process in an irreversible way, when the system is
unable to reach true thermal equilibrium, as is the case in aerogels.
On the other hand, if small steps are used, measurements errors
accumulate, resulting in a large error on the adsorbed amount.  In our
previous work, we instead used a continuous method, in which helium is
injected at a regulated flowrate, small enough (typically 0.1 STP
cc/min) to minimize the thermal perturbation.  Due to this small value
(2\% of the range of our regulated flowmeter, Brooks 5850S), the
flowrate had to be measured by deriving the flow to a calibration
reservoir every time it was modified.  Even so, it was very difficult
to precisely measure hysteresis loops.  Going from filling to
emptying, or reversely, involves the manipulation of valves, and
associated errors in the amount of helium in the experimental cell,
which accumulate over repeated cycles.  This precludes any precise
determination of the temperature at which the hysteresis loop
disappears.

A solution would be to use the method of Herman \it{et al.}\rm
\cite{HermanPRB05}.  These authors directly measure the adsorbed
amount in the gel by a capacitive technique.  This however requires to
plate the porous medium with electrodes.  Here, we present a more
general alternative, where the adsorbed amount is controlled through
the temperature of a helium gas reservoir connected to the
experimental cell.  For a given amount of helium in the whole system,
varying the reservoir temperature transfers atoms from the reservoir
to the cell, or reversely.  For a given temperature, the amount of
adsorbed helium will be the same, independently of the previous
history.  For large enough pressures, this provides an efficient way
to control this amount.
\begin{figure}
   \centerline{\includegraphics[height=2.in]{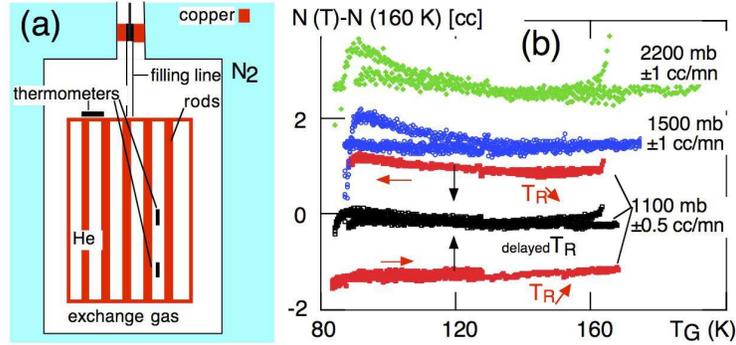}}
 \caption[boite]{(a) Sketch of the reservoir.  (b) Temperature 
 dependance of the computed total 
 amount $N_{T}$ of helium (in STP cm$^{3}$).  The curves correspond to three
 loading pressures at 160~K. For the 1100~mb data, $N_{T}$ computed
 taking the gas temperature $T_{G}$ equal to the reservoir temperature
 $T_{R}$ presents a $\pm$~1 STP cm$^{3}$ hysteresis between filling
 and emptying.  This hysteresis is suppressed when assuming a 60~s
 thermal lag between $T_{G}$ and $T_{R}$.  The same lag suppresses the
 hysteresis for the 1500 and 2200~mb data as well (for clarity, the
 corresponding curves are vertically shifted).  }
  \label{boite}
  \end{figure} 
 \section{Constraints and design}\label{constraints}
The design parameters of the system are set by a number of
constraints.  First, the reservoir volume and the temperature range of
operation are fixed by the amount to transfer.  Second, the reservoir
design should be such that the gas temperature remains homogeneous when
the heating power is varied, i.e. the static and dynamic gradients
should be as small as possible.  Third, the coupling of the reservoir
to the cold source should be good enough to allow cooling at the
required rate (heating being not a problem).

We use a cylindrical copper reservoir (10 cm long, 6.5~cm diameter,
with 2.5~mm thick walls, and approximate volume 230~cm$^{3}$), inside
a vacuum can immersed in a tank of liquid nitrogen
(Fig.\ref{boite}(a)).  The 1$\times$ 1.5~mm filling line has a volume
of less than 1 cm$^{3}$.  At a constant pressure of $P$ bar, varying
the reservoir temperature from 80~K to 180~K results in a transfer of
about 400 $P$ STP cm$^{3}$ .  This is more than needed to cover the
steep part of the isotherm for $T>4.2$~K in a 0.5~cm$^{3}$ aerogel
(Fig.\ref{hyster}).  There are three paths for the thermal contact to
the nitrogen bath : conduction through the 3$\times$4~mm stainless
steel tube to which the reservoir is suspended, thermal radiation to
the walls of the vacuum can, and conduction through the exchange gas
(air).  This latter contribution dominates by a factor of
order 20 in our conditions, where the exchange gas pressure is of
order 50~Pa, at the crossover between the molecular and hydrodynamic
regimes of heat transfer.  This choice allows a good thermal coupling,
with a nearly uniform cooling power on the external surface of the
reservoir.  Heating power is provided by a 60~ohms manganin wire
heater wound on the external faces of the reservoir, in such a way
that the area enclosed between two turns is approximately constant.
We expect this design to ensure an uniform temperature of the copper.
In order to maximize the thermal coupling to the gas, the reservoir is
also traversed by twenty-four 4~mm diameter copper rods, soldered
about 12~mm apart on its top and bottom faces.  A platinum resistance,
glued on the top face of the reservoir, is used for regulating its
temperature, while two platinum resistances inside the reservoir read
the gas temperature between two rods (at mid-height and close to the
bottom).

The reservoir temperature is regulated with a maximal heating power of
10~W, allowing to go from 80~K up to 160~K in one hour.  The minimal
time (heater off) to come back to 80~K is of order of four hours.  The
maximum flow rate which can be obtained over the whole temperature
range is 2~cm$^{3}$/minute for filling and 0.5~cm$^{3}$/minute for
emptying.  These values are more than enough for studying aerogels.
\section{Accuracy}\label{precision}
We first studied the accuracy of the system by coupling it to a
temperature regulated reference reservoir at 305~K. For a given total
amount of helium, the pressure of the whole system was measured as a
function of the cold reservoir temperature, from which we computed the
amounts $N_{C}$ and $N_{W}$ of helium in the cold reservoir and the
remaining of the system (reference reservoir, pressure gauge, and
connecting lines at regulated ambiant temperature, plus filling line
to the reservoir).  Experimental volumes were determined at ambiant
temperature by expanding gas from the reference volume to the other
parts of the set-up, and measuring the pressure changes. We
calibrated our thermometers from one of our experimental runs, using
our cold reservoir as a gas thermometer.  This gives the correct
boiling nitrogen temperature within 0.1~K. For other runs,
$N_{C}+N_{W}$ has to be constant over the whole temperature range,
independently of the rate of variation of temperature (which is set by
the requested flowrate) or of the coupling to the cold source (i.e. of
the exchange gas pressure).  Results for a loading pressure of
1100~mbar at 160~K are shown in Fig.~\ref{boite}(b) for filling and
emptying at $\pm$ 0.5 cm$^{3}$/minute.  The clear 2~cm$^{3}$
hysteresis is much larger than what could be expected from the
impedance of the filling line\cite{impedance}, or from the possible
changes in its temperature profile.  It results from a delay between
the reservoir temperature $T_{R}$ and the gas temperature $T_{G}$,
evidenced by the inner thermometers.  Their measured delay time (65 to
75~s) is much larger than computed from the thermal diffusivity of the
gas and the inter-rods distance (between 0.2 and 1~s depending on
temperature and pressure).  This indicates some unexpected thermal
resistance within the copper, or between the copper and the gas.
Anyway, this delay time can be efficiently corrected for.  Using
$T_{G}= T_{R}-\tau \dot{T_{R}}$ with a delay time $\tau$=60~s to
compute $N_{C}$ suppresses most of the hysteresis for all temperature,
pressure, and flow rates studied, as illustrated by the corrected
curves in Fig.~\ref{boite}(b).  All together, the precision and
reproducibility are better than one STP cm$^{3}$, on a total of 200 to
350 transferred STP cm$^{3}$ (depending on pressure), for flow rates
up to 2 cm$^{3}$/minute.  Dynamic temperature gradients are thus not a
problem.  Static gradients are not a problem either : when the
exchange gas is pumped at a constant reservoir temperature of 160~K,
which results in a decrease of the heating power from 6~W to 0.3~W,
the calculated total amount changes by less than 1 STP cm$^{3}$.
\section{Application to aerogels}\label{aerogel}
\begin{figure}
   \centerline{\includegraphics[height=1.8in]{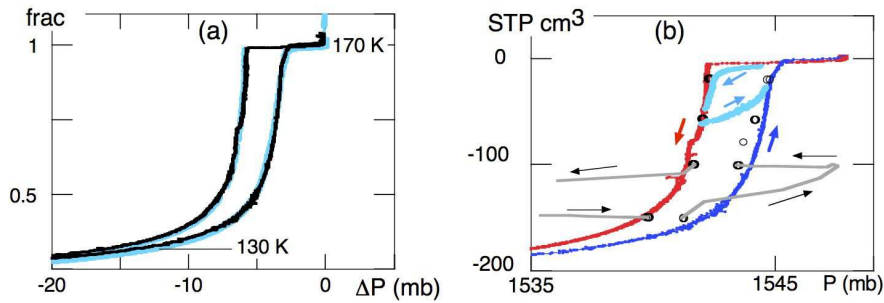}}
 \caption[hyster]{(a) Adsorption isotherm at 4.7~K for aerogel N102
 \cite{Lambert}.  Condensed fraction as a function of $\Delta P$, the
 pressure difference with the saturation pressure, for the new (black)
 and the previous (grey) data.  The reservoir temperature is indicated
 for two points; (b) Comparison of continuous filling/emptying and of
 finite steps (open dots).  The adsorbed amount is refered to the
 filled aerogel and expressed in STP cm$^{3}$.  Some transients are
 shown (grey lines).  Two scanning curves are also shown.}
 \label{hyster}
  \end{figure}

We used this device to perform adsorption and desorption isotherms at
4.7~K on a neutrally catalyzed aerogel\cite{Lambert}.
Figure~\ref{hyster}a shows that the inferred condensed fraction
compares well to our former results \cite{Lambert}.  The remarkable
point is that the perfect closing of the loop at the lower pressures
is now much easier to obtain.

With the new system, the temperature control offers the possibility to
easily perform complex paths in the (pressure-adsorbed amount) plane.
This is illustrated by Fig~\ref{hyster}b.  The lines internal to the
main hysteresis loop are obtained by reversing from filling to
emptying , or reversely, while being on the hysteretic part of the
adsorption loop (they are the so-called scanning curves).  These
operations, which are delicate when using the flowmeter, become
trivial here.  Also, it is easy to fill or empty in steps of finite
size (grey lines), by changing the temperature set point by steps.  As
we explained in the introduction, this strongly perturbs the system,
probably reflecting the change of the saturated vapor pressure due to
local heating or cooling.  The interesting fact is that, once the
change is relaxed, the final state on the desorption branch coincides
with that obtained for slow emptying (0.07 STP cm$^{3}$/minute).
There is a difference on the adsorption branch, but future work is
needed to know whether this is due to the larger flowrate used for
continuous filling (0.2 STP cm$^{3}$/minute).  In any case, the
important thing is that the temperature control is well adapted for
studying the relaxation processes in adsorption and desorption.

In conclusion, the set-up we have developed offers an easy and 
precise control of the adsorbed amount in a porous system. It is thus 
ideal to study hysteresis, relaxation, and metastable states in the 
adsorption and desorption of fluids in porous media.


\begin{thebibliography}{99}
\bibitem{Gelb99}
For a review, see  L. D. Gelb {\it et al.}, {\it  Rep. Prog. Phys.} 
{\bf 62}, 1573 (1999). 

\bibitem{Wong90}
 A. P. Y. Wong and M. H.W. Chan, {\it Phys. Rev. Lett.}  {\bf65}, 2567 (1990). 

\bibitem{Tulimieri99}
 D. J. Tulimieri, J. Yoon, and M. H. W. Chan, {\it Phys. Rev. Lett.}
{\bf 82}, 121 (1999).

\bibitem{Gabay00}
C. Gabay {\it et al.}, {\it J. Low Temp. Phys.}  {\bf 121}, 585 (2000).

\bibitem{HermanTout}
 J. Beamish and T. Herman, {\it Physica B}  {\bf340}, 329 (2003)
; J. Beamish and T. Herman, {\it J. Low Temp. Phys. } {\bf134}, 339 (2004).

\bibitem{Lambert04}
T. Lambert {\it et al.}, {\it J. Low Temp. Phys.}  {\bf 134}, 293 (2004).

\bibitem{HermanPRB05}
T. Herman, J. Day, and J. Beamish, {\it Phys. Rev. B}  {\bf72}, 184202 (2005).

\bibitem{Kierlik01}
E. Kierlik {\it et al.}, {\it Phys. Rev. Lett.}  {\bf 87}, 055701 (2001).

\bibitem{Detcheverry05}
F. Detcheverry, E. Kierlik, M. L. Rosinberg, and G. Tarjus, {\it 
Phys. Rev. E.}  {\bf 68}, 61504 (2003).
F. Detcheverry, E. Kierlik, M. L. Rosinberg, and G. Tarjus, {\it 
Langmuir}  {\bf 20}, 8006 (2004).
F. Detcheverry, E. Kierlik, M. L. Rosinberg, and G. Tarjus, {\it 
Phys. Rev. E.}  {\bf 72}, 051506 (2005).

\bibitem{impedance}
The pressure loss is 
less than 10 Pa for a flowrate of 1~cm$^3$/s, corresponding to a fraction  of cm$^3$
difference  in $N_{C}$ between filling and emptying .

\bibitem{Lambert}
T. Lambert, L. Puech, and P.E. Wolf, {\it cond-mat}/0603486.
\end{thebibliography}
\end{document}